\documentclass[fleqn,usenatbib]{mnras}

\usepackage{newtxtext,newtxmath}
\usepackage[T1]{fontenc}

\DeclareRobustCommand{\VAN}[3]{#2}
\let\VANthebibliography\thebibliography
\def\thebibliography{\DeclareRobustCommand{\VAN}[3]{##3}\VANthebibliography}

\usepackage{graphicx}	
\usepackage{amsmath}	

\title[Large-amplitude Filament Oscillations]{Large-amplitude Oscillations of a Quiescent Filament Excited by an Extreme-ultraviolet Wave}

\author[W. W. Pan et al.]{
W. W. Pan,$^{1,2}$
Q. M. Zhang,$^{1,3}$\thanks{E-mail: zhangqm@pmo.ac.cn}
and Y. Qiu$^{4}$
\\
$^{1}$Key Laboratory of Dark Matter and Space Astronomy, Purple Mountain Observatory, Nanjing 210023, China\\
$^{2}$School of Astronomy and Space Science, University of Science and Technology of China, Hefei 230026, China\\
$^{3}$State Key Laboratory of Solar Activity and Space Weather, Beijing 100190, China\\
$^{4}$Institute of Science and Technology for Deep Space Exploration, Suzhou Campus, Nanjing University, Suzhou 215163, China
}

\date{Accepted XXX. Received YYY; in original form ZZZ}

\pubyear{2025}

\begin{document}
\label{firstpage}
\pagerange{\pageref{firstpage}--\pageref{lastpage}}
\maketitle

\begin{abstract}
In this paper, we carry out multiwavelength observations of simultaneous longitudinal and transverse oscillations of a quiescent filament 
excited by an extreme-ultraviolet (EUV) wave on 2023 February 17.
A hot channel eruption generates an X2.3 class flare and a fast coronal mass ejection (CME) in active region (AR) NOAA 13229 close to the eastern limb.
The CME drives an EUV wave, which propagates westward at a speed of $\sim$459 km s$^{-1}$.
After arriving at the filament $\sim$340.3 Mm away from the flare site, 
the filament is disturbed and starts large-amplitude oscillations, which are mainly observed in 171 {\AA}.
The longitudinal oscillations last for nearly two cycles. The average initial amplitude, velocity, period, and damping time are 
$\sim$4.7 Mm, $\sim$26.5 km s$^{-1}$, $\sim$1099.1 s, and $\sim$2760.3 s, respectively.
According to the pendulum model, the curvature radius and minimum horizontal magnetic field strength of the dips are estimated to be 6.7$-$9.9 Mm and 4.6$-$5.6 G.
The transverse oscillations last for 2$-$3 cycles. The average initial amplitude, velocity, period, and damping time are
$\sim$1.8 Mm, $\sim$11.2 km s$^{-1}$, $\sim$994.4 s, and $\sim$3576.2 s, respectively. The radial magnetic field strength of the dips are estimated to be 6.6$-$7.4 G.
\end{abstract}




\section{Introduction} \label{sect:intro}
Filaments are dark magnetic structures on the solar disk when observed in H$\alpha$ wavelength \citep{mac10,lab10}.
In general, they are $\sim$10$^{2}$ times cooler and denser than the corona \citep{par14,lab22}.
The plasmas in filaments are supported by magnetic tension force of magnetic dips 
in sheared arcades or flux ropes \citep{pri89,luna12,luna17,zhou17,gib18,guo23,chen25,yan25a}.
Filaments with dextral (sinistral) chirality are dominant in northern (southern) hemisphere \citep{mar98,ou17}.
According to the positions, filaments are divided into three categories: active region (AR) filaments, intermediate filaments, and quiescent filaments \citep{yan25b}.

Although filaments are stable for a few days to several weeks, they are sensitive to external disturbances. 
A filament may experience small-amplitude or large-amplitude oscillations as a result of perturbations, such as flares, microflares, coronal jets, and extreme-ultraviolet (EUV) waves \citep{oli02,arr18,tri09}. 
Small-amplitude oscillations (SAOs) are prevailing with periods of a few minutes and velocity amplitudes of $\leq$10 km s$^{-1}$ \citep{oka07,hil13,li18,song24}.
Large-amplitude oscillations (LAOs) have velocity amplitudes of $\ge$10 km s$^{-1}$, which usually damp with time as a result of energy dissipation \citep{luna18}.
Like coronal loop oscillations \citep{god16}, the damping times ($\tau$) decrease with increasing amplitudes \citep{zqm13}.
According to the direction of oscillations, LAOs are further divided into longitudinal oscillations (LALOs) and transverse oscillations (LATOs).
The dense materials move back and forth along the field lines in LALOs \citep{jing03,jing06,vrs07,li12,lk12,zqm12,zqm17b,zqm20}, 
while the whole body swings perpendicular to the spine of the filament in LATOs \citep{kle69,eto02,her11,asai12,dai12,liu13,zyj24}.
Sometimes, a filament may experience both longitudinal and transverse oscillations \citep{pant16,zqm17a,dai21}.
The vertically oscillating filaments are also named winking filaments in that the filaments show up 
in the red wing and blue wing of H$\alpha$ waveband alternatively and last for several cycles before fading out \citep{hyd66,ram66,shen14a,dai23}.
Occasionally, LAOs precede filament eruptions, indicating that filament oscillations may stand for a meta equilibrium and serve as 
one of the precursors for solar eruptions \citep{iso06,chen08,li12,bi14,kol16,dai21,zqm20,zqm24,bec24,li25}.

For LALOs, the triggering mechanisms are diverse, including microflares (subflares) \citep{jing03,vrs07,zqm12}, flares \citep{li12,zqm20,luna24}, 
coronal jets \citep{luna14,ni22,jos23,tan23}, and EUV waves \citep{shen14b}. 
The dominant restoring force is believed to be the gravity of filament threads along the dips \citep{lk12,zqm12,zqm13,zhou14}.
Therefore, LALOs could be well described by a pendulum model, in which the period of oscillation ($P_L$) is exclusively dependent on the curvature radius ($R$) of the dip:
\begin{equation} \label{eqn-1}
P_L=2\pi\sqrt{\frac{R}{g_{\odot}}},
\end{equation}
where $g_{\odot}=274$ m s$^{-2}$ denotes the gravitational acceleration at the photosphere \citep{lk12,luna16}.
Moreover, the minimum horizontal magnetic field strength of the dip could be determined.
The main damping mechanisms of LALOs include mass accretion \citep{rud16}, radiative loss, thermal conduction \citep{zqm13}, and wave leakage \citep{zly19}.

For LATOs, the triggering mechanisms are coronal jets \citep{zqm17a}, Moreton waves, and EUV waves \citep{eto02,gil08,her11,liu13}.
\citet{shen14a} proposed a cartoon model to show the interaction between a shock wave and a filament. 
LATOs and LALOs would be excited when the direction of the shock wave is exactly perpendicular and parallel to the filament axis, respectively.
Both types of oscillations are expected to coexist when the direction of shock wave is oblique. However, coexistence of LATOs and LALOs has been rarely reported till now.
\citet{pant16} investigated an eruptive M-class flare associated with a coronal mass ejection (CME), which drives a shock wave on 2013 March 15.
Simultaneous longitudinal and transverse oscillations of a remote AR filament are excited by the shock wave. 
Magnetic field strength and mass accretion rate of the filament are estimated.
\citet{dai21} studied a sympathetic eruption of a large quiescent filament after the primary eruption of a small filament on 2015 April 28.
Before the eruption of the large filament, longitudinal oscillations with an initial speed of $\sim$43 km s$^{-1}$ 
and transverse oscillations with an initial speed of $\sim$14 km s$^{-1}$ are detected. 
Continuous mass drainage at one footpoint of the filament is found to accompany with LAOs, which is indicative of a gradual process of loss of equilibrium.
In this paper, we report another case study of simultaneous LALOs and LATOs of a quiescent filament induced by an incoming EUV wave 
as a result of an eruptive X2.3 flare originating from AR NOAA 13229 (N25E67) on 2023 February 17. 
The paper is organized as follows. In Section~\ref{sect:obs}, we describe observations and data analysis.
The results are presented in Section~\ref{sect:res}. We compare our results with previous works in Section~\ref{sect:dis} and draw a conclusion in Section~\ref{sect:sum}.

\section{observations and data analysis} \label{sect:obs}
\begin{figure}
    \centering
    \includegraphics[width=\columnwidth]{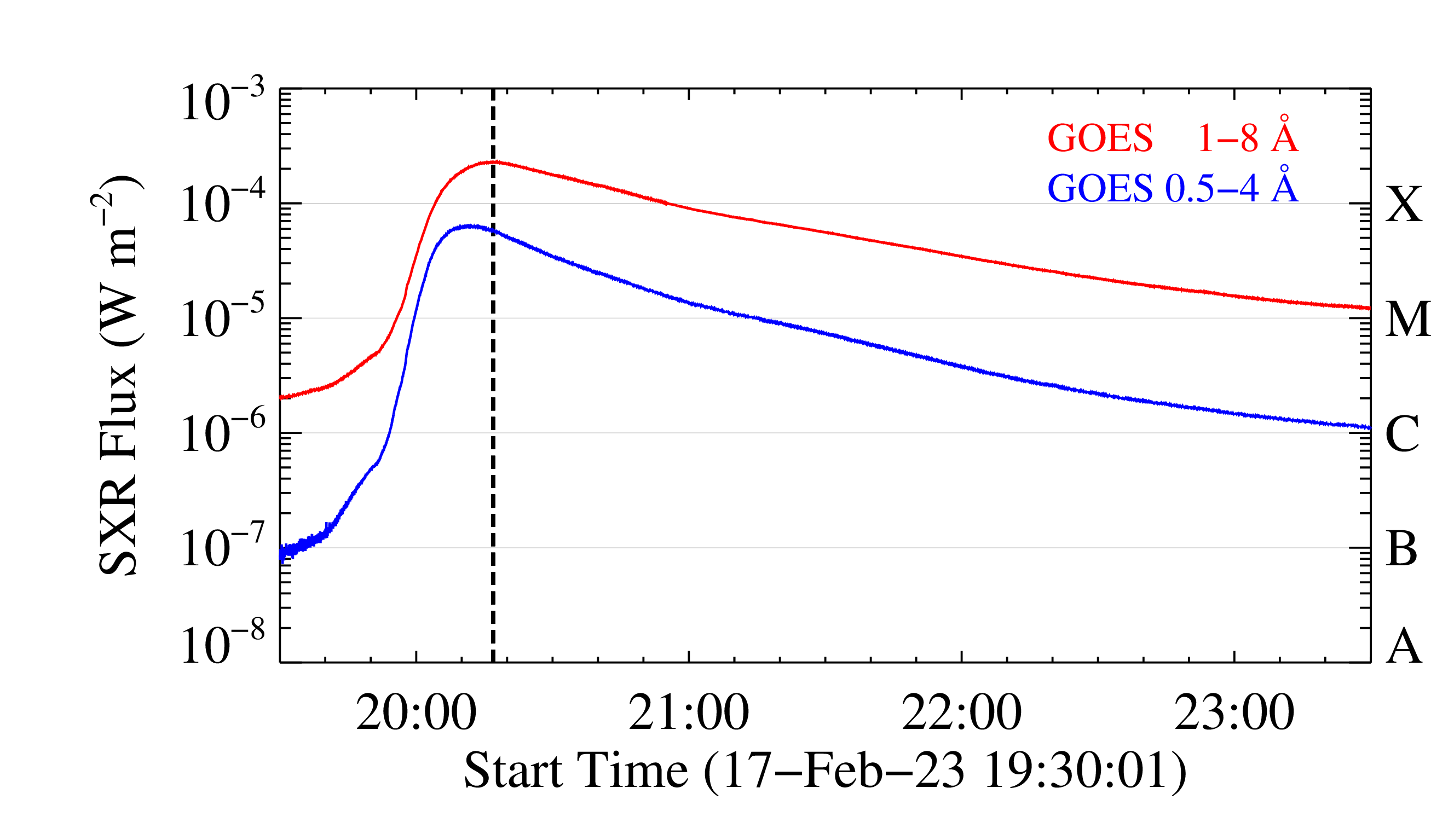}
    \caption{SXR fluxes of the X2.3 class flare in 0.5$-$4 {\AA} (blue line) and 1$-$8 {\AA} (red line). 
    The flare peak time (20:16 UT) is marked with a vertical dashed line.}
    \label{fig1}
\end{figure}

The X2.3 class flare and quiescent filament were observed by the H$\alpha$ Imaging Spectrograph \citep[HIS;][]{qiu22} 
on board the Chinese H$\alpha$ Solar Explorer \citep[CHASE;][]{li22} spacecraft, 
the Solar Upper Transition Region Imager \citep[SUTRI;][]{bai23} on board the Space Advanced Technology demonstration satellite (SATech-01),
and the Atmospheric Imaging Assembly \citep[AIA;][]{lem12} on board the Solar Dynamics Observatory \citep[SDO;][]{pes12}.
CHASE/HIS has a spatial resolution of 2$\arcsec$ and a time cadence of $\sim$60 s.
SUTRI takes full-disk images in Ne {\sc vii} 465 {\AA} ($T\sim0.5$ MK) with a spatial resolution of 2.46$\arcsec$ and a time cadence of $\sim$30 s \citep{tian17}.
SDO/AIA takes full-disk images in two ultraviolet (UV; 1600 {\AA} and 1700 {\AA}) and seven extreme-ultraviolet (EUV; 94, 131, 171, 193, 211, 304, and 335 {\AA}) wavelengths.
The EUV passbands have a spatial resolution of 1.2$\arcsec$ and a time cadence of 12 s.
Full-disk line-of-sight (LOS) magnetograms of the photosphere were observed by the Helioseismic and Magnetic Imager \citep[HMI;][]{sch12} on board SDO.
The magnetograms have a spatial resolution of 1.2$\arcsec$ and a time cadence of 45 s.
The level\_1 data of AIA and HMI are calibrated by using the standard Solar Software (SSW) programs \texttt{aia\_prep.pro} and \texttt{hmi\_prep.pro}, respectively.
The H$\alpha$ and 465 {\AA} images are coaligned with AIA 304 {\AA} images with an accuracy of 2$\arcsec$.
The associated full-halo CME\footnote{https://cdaw.gsfc.nasa.gov/CME\_list/} was observed by the C2 and C3 of 
the Large Angle and Spectrometric Coronagraph \citep[LASCO;][]{bru95} on board the Solar and Heliospheric Observatory (SOHO) spacecraft.
Soft X-ray (SXR) fluxes of the flare in 0.5$-$4 {\AA} and 1$-$8 {\AA} were recorded by the Geostationary Operational Environmental Satellite (GOES) spacecraft 
with a cadence of $\sim$2 s and plotted in Figure~\ref{fig1} with blue and red lines, respectively.
The fluxes start to rise at $\sim$19:38 UT and reach the peak values at $\sim$20:16 UT (vertical dashed line), which is followed by a gradual decay.
The radio dynamic spectra in the 45$-$90 MHz frequency range during the flare were obtained from the MEXICO-LANCE-A station 
belonging to the e-CALLISTO\footnote{https://www.e-callisto.org/} network.

\begin{figure*}
    \centering
    \includegraphics[width=1.8\columnwidth]{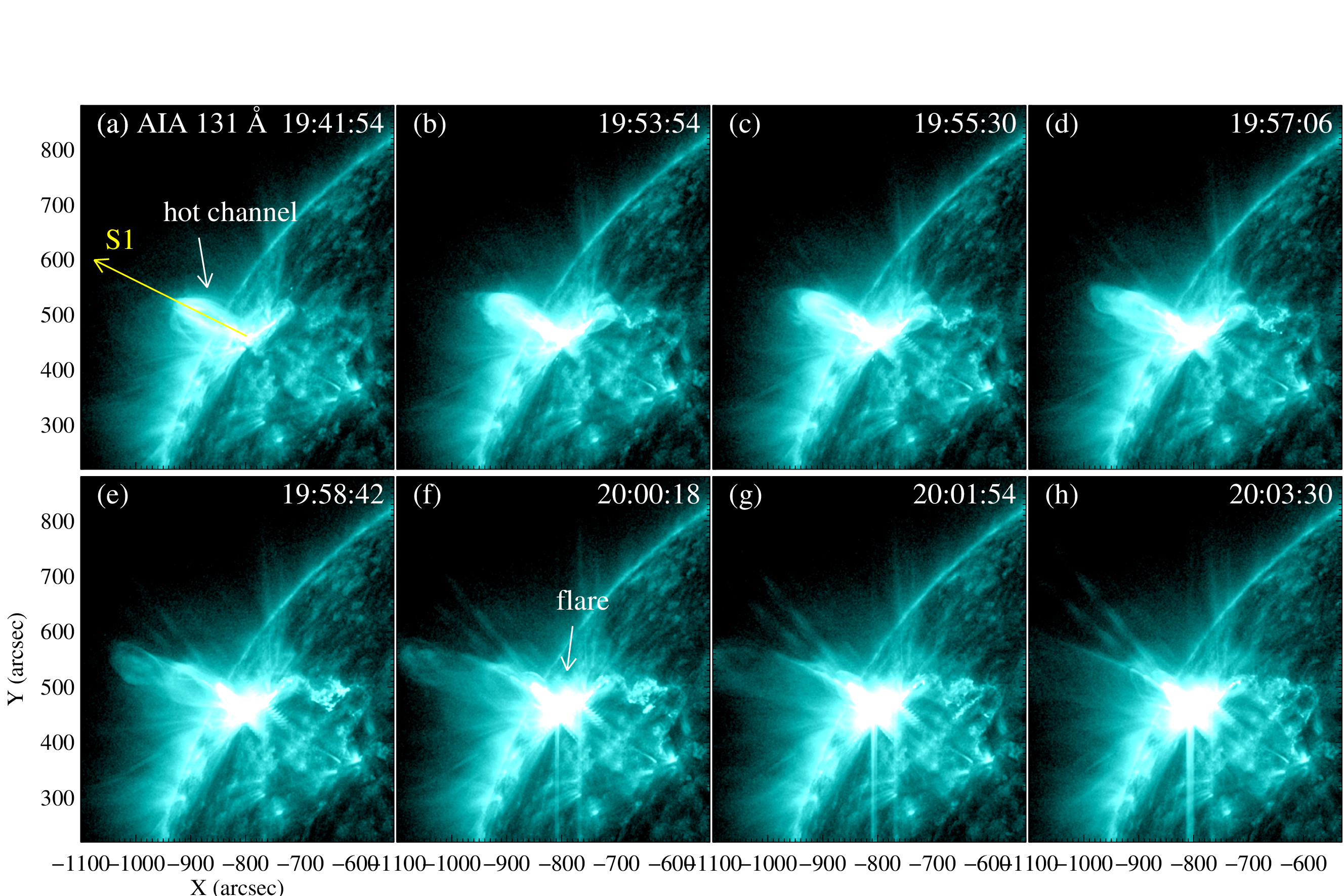}
    \caption{AIA 131 {\AA} images, showing the evolution of HC eruption and flare, which are pointed by the white arrows.
    In panel (a), the yellow arrow illustrates the straight slice (S1) with a length of $\sim$222 Mm.
    An online animation of AIA 131 {\AA} images is available as Supplementary material. 
    The $\sim$7 s animation covers from 19:41 to 20:03 UT.}
    \label{fig2}
\end{figure*}

\section{results} \label{sect:res}

\subsection{Hot channel eruption and X-class flare} \label{hc}
\begin{figure}
    \centering
    \includegraphics[width=0.9\columnwidth]{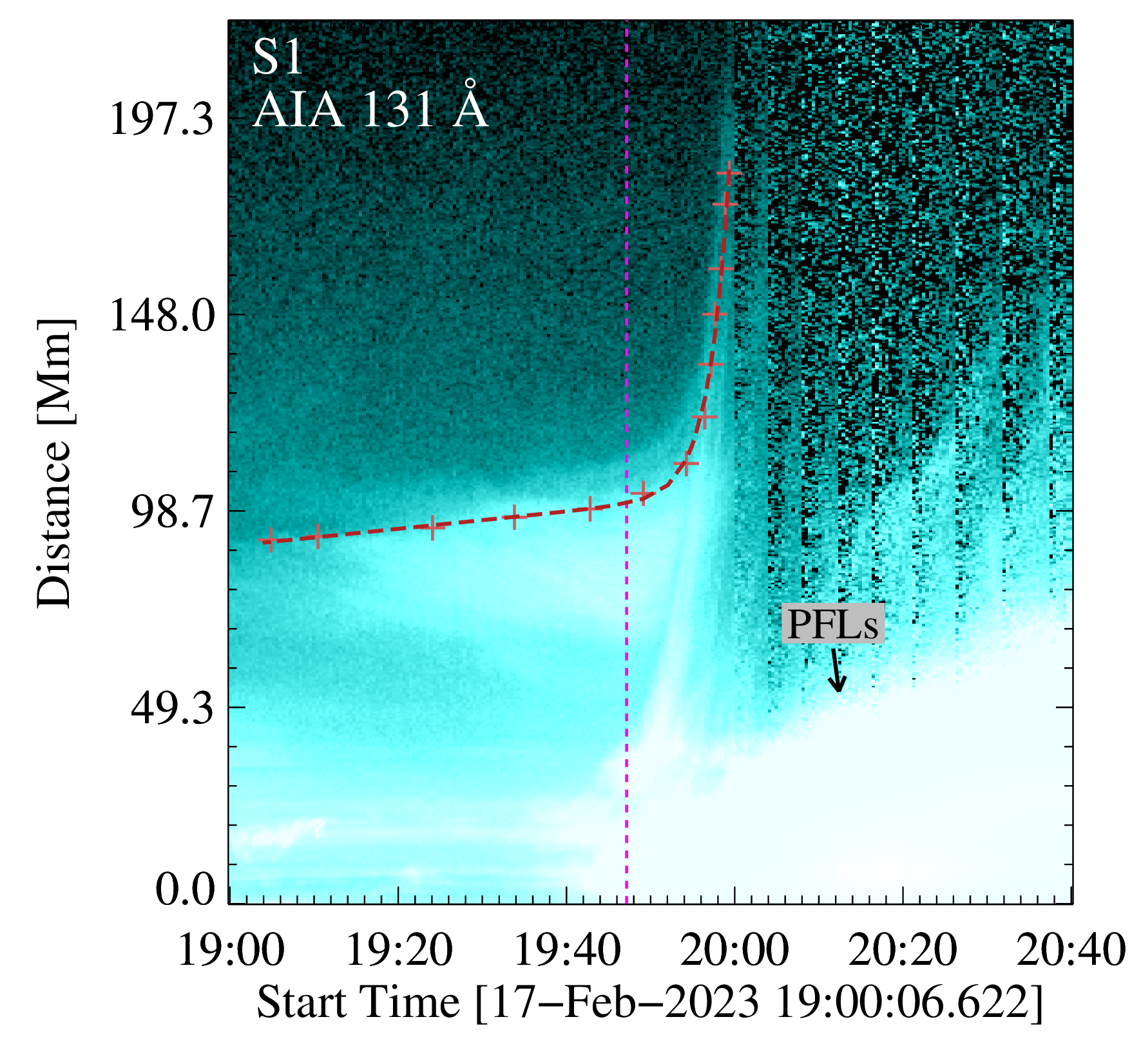}
    \caption{Time-distance diagram of S1 in AIA 131 {\AA} during 19:00$-$20:40 UT. 
    $s=0$ and $s=222$ Mm denote the southwest and northeast endpoints of S1, respectively.
    The red plus symbols stand for the trajectory of the HC along S1. 
    The result of curve fitting using Equation~\ref{eqn-2} is superposed with a red dashed line.
    The magenta dashed line denotes the onset time (19:47 UT) of fast rise of the HC.
    The black arrow points to the hot post-flare loops (PFLs).}
    \label{fig3}
\end{figure}

Figure~\ref{fig2} shows eight snapshots of AIA 131 {\AA} images during 19:40$-$20:03 UT (see also the online animation \textit{anim1.mp4}).
In panel (a), a hot channel (HC) erupts from AR 13229. A HC is believed to be a flux rope with extremely high temperature ($T\sim10$ MK), 
which is predominantly observed in AIA 94 and 131 {\AA} \citep{cx11,cx12,li16,zqm22,zqm23,zqm24}.
The HC propagates in the northeast direction, generating the X2.3 flare underneath (see panel (f)). 
To investigate the kinematics of the HC, we select a straight slice (S1) with a length of $\sim$222 Mm along the direction of propagation in panel (a).
Time-distance diagram of S1 in 131 {\AA} during 19:00$-$20:40 UT is plotted in Figure~\ref{fig3}.
The trajectory of HC is marked with red plus symbols, featuring a slow rise phase and a fast rise phase during 19:00$-$20:00 UT 
before escaping the field of view (FOV) of AIA.
To perform a curve fitting of the trajectory, we use the following function \citep{cx13}:
\begin{equation} \label{eqn-2}
 	h(t)=c_0 e^{(t-t_0)/ \sigma} +c_1(t-t_0) + c_2,
\end{equation}
where $t$ is time, $t_0$ is set to be 19:04 UT, $h(t)$ is the height along S1,  
$\sigma$, $c_0$, $c_1$, and $c_2$ are free parameters. The onset time of fast rise phase is defined as:
\begin{equation} \label{eqn-3}
 	t_{\mathrm{onset}} =\sigma \ln(c_1 \sigma /c_0) +t_0.
\end{equation}
The result of curve fitting using Equation~\ref{eqn-2} is superposed with a red dashed line in Figure~\ref{fig3}, showing that the fitting is quite good \citep{zqm22,zqm23}.
$t_{\mathrm{onset}}$ ($\sim$19:47 UT) is earlier than the flare peak time at 20:16 UT.

\begin{figure}
    \centering
    \includegraphics[width=\columnwidth]{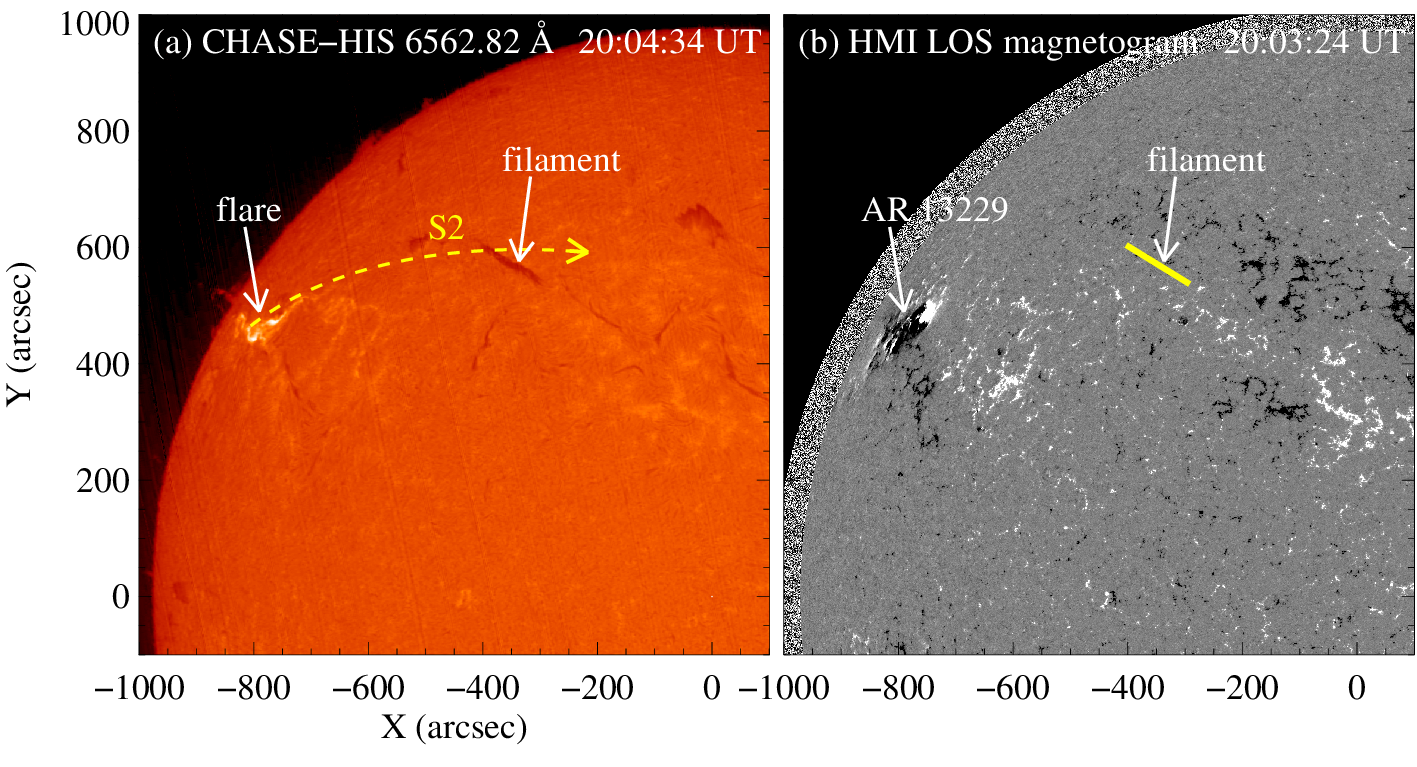}
    \caption{(a) CHASE/HIS H$\alpha$ image at 20:04:34 UT. The white arrows point to the flare and quiescent filament.
    The yellow dashed line (S2) is used to investigate the EUV wave, which impacts the filament.
    (b) SDO/HMI LOS magnetogram at 20:03:24 UT. The white arrows point to AR 13229 and position of the filament (short yellow line).}
    \label{fig4}
\end{figure}

In Figure~\ref{fig4}, the left panel shows CHASE/HIS H$\alpha$ image at 20:04:34 UT. The X2.3 flare and quiescent filament are indicated by white arrows.
The distance between the flare site and filament is $\sim$340.3 Mm. The right panel shows SDO/HMI LOS magnetogram at 20:03:24 UT. 
The arrows point to AR 13229 hosting the flare and position of the filament (short yellow line) along the polarity inversion line (PIL).

\begin{figure*}
    \centering
    \includegraphics[width=1.5\columnwidth]{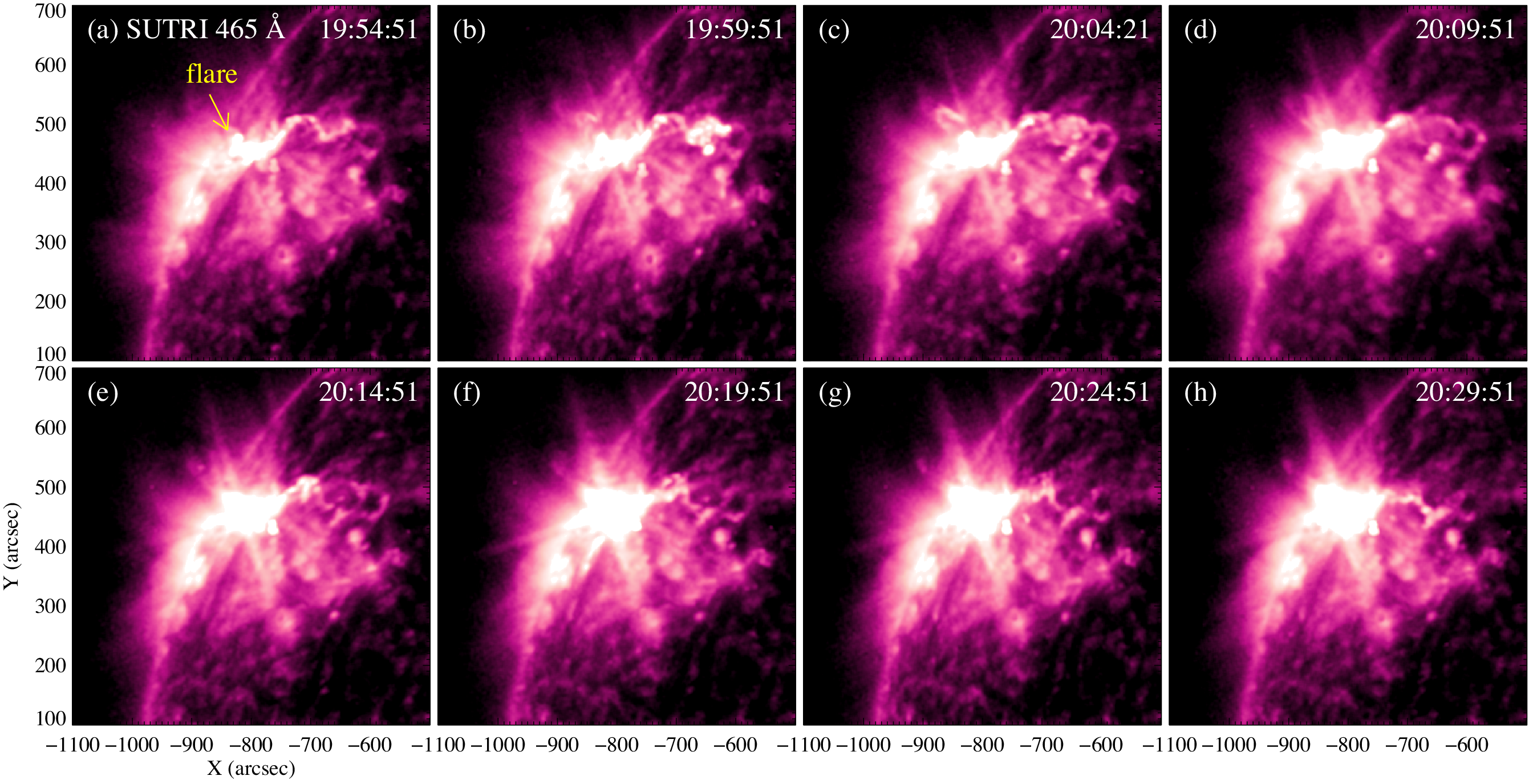}
    \caption{SUTRI 465 {\AA} images, showing the X2.3 flare during 19:54$-$20:29 UT in AR 13229.
    An online animation of 465 {\AA} images is available as Supplementary material. 
    The $\sim$5 s animation covers from 19:54 to 20:47 UT.}
    \label{fig5}
\end{figure*}

Fortunately, the flare was also detected by SUTRI during 19:54$-$20:47 UT.
In Figure~\ref{fig5}, the eight snapshots of 465 {\AA} images illustrate the evolution of the X2.3 flare (see also the online animation \textit{anim2.mp4}).
The total emission of the flare increases from 19:54 UT and reaches the peak value at 20:20:30 UT before declining gradually.
Therefore, the peak time in 465 {\AA} is $\sim$4.5 minutes later than that in 1$-$8 {\AA}, suggesting a cooling process of the post-flare loops \citep{car94,zqm19}.

\subsection{CME and shock wave} \label{cme}
\begin{figure*}
    \centering
    \includegraphics[width=1.5\columnwidth]{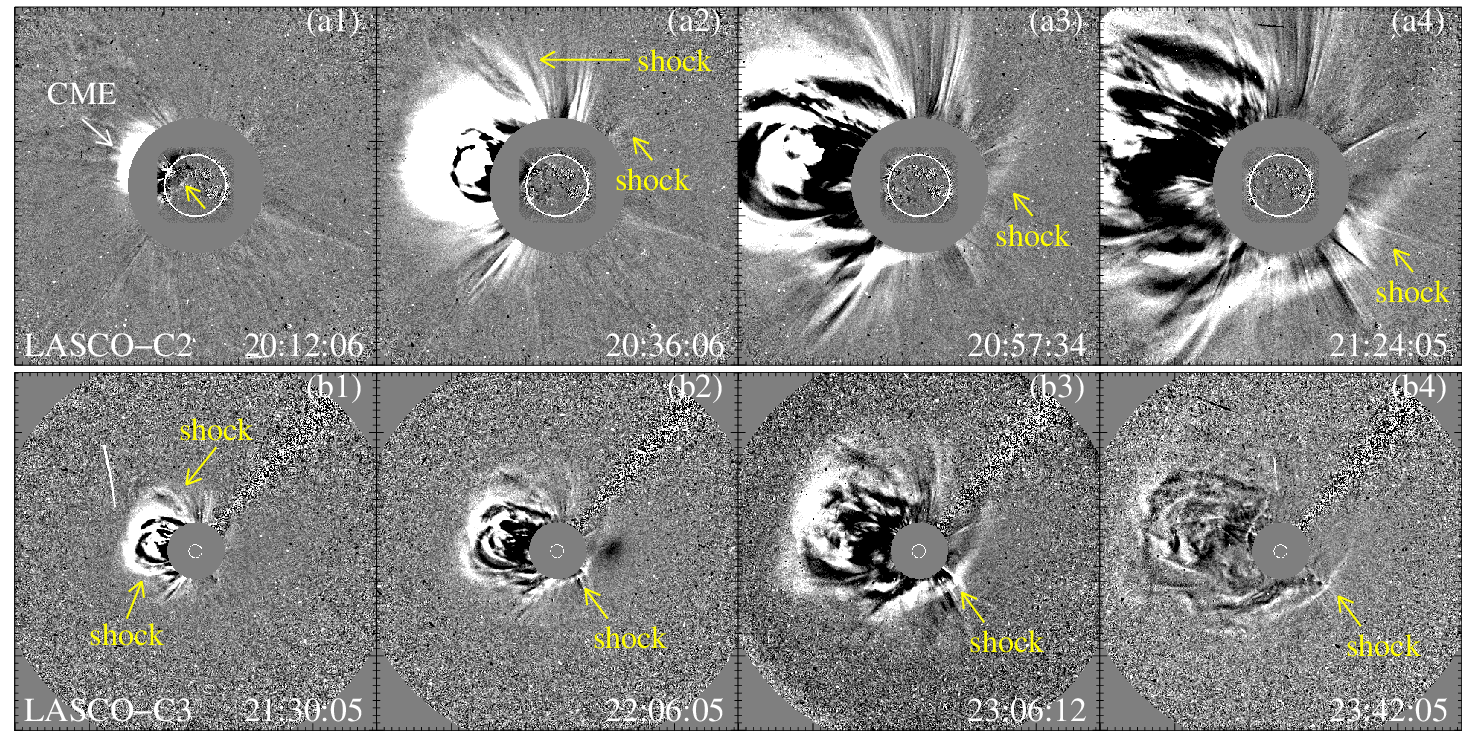}
    \caption{Top panels: Running-difference images of the CME observed by LASCO-C2 during 20:12$-$21:24 UT.
    Bottom panels: Running-difference images of the same CME observed by LASCO-C3 during 21:30$-$23:42 UT. 
    The white arrow points to the CME leading edge. The yellow arrows point to the CME-driven shock.}
    \label{fig6}
\end{figure*}

\begin{figure}
    \centering
    \includegraphics[width=0.9\columnwidth]{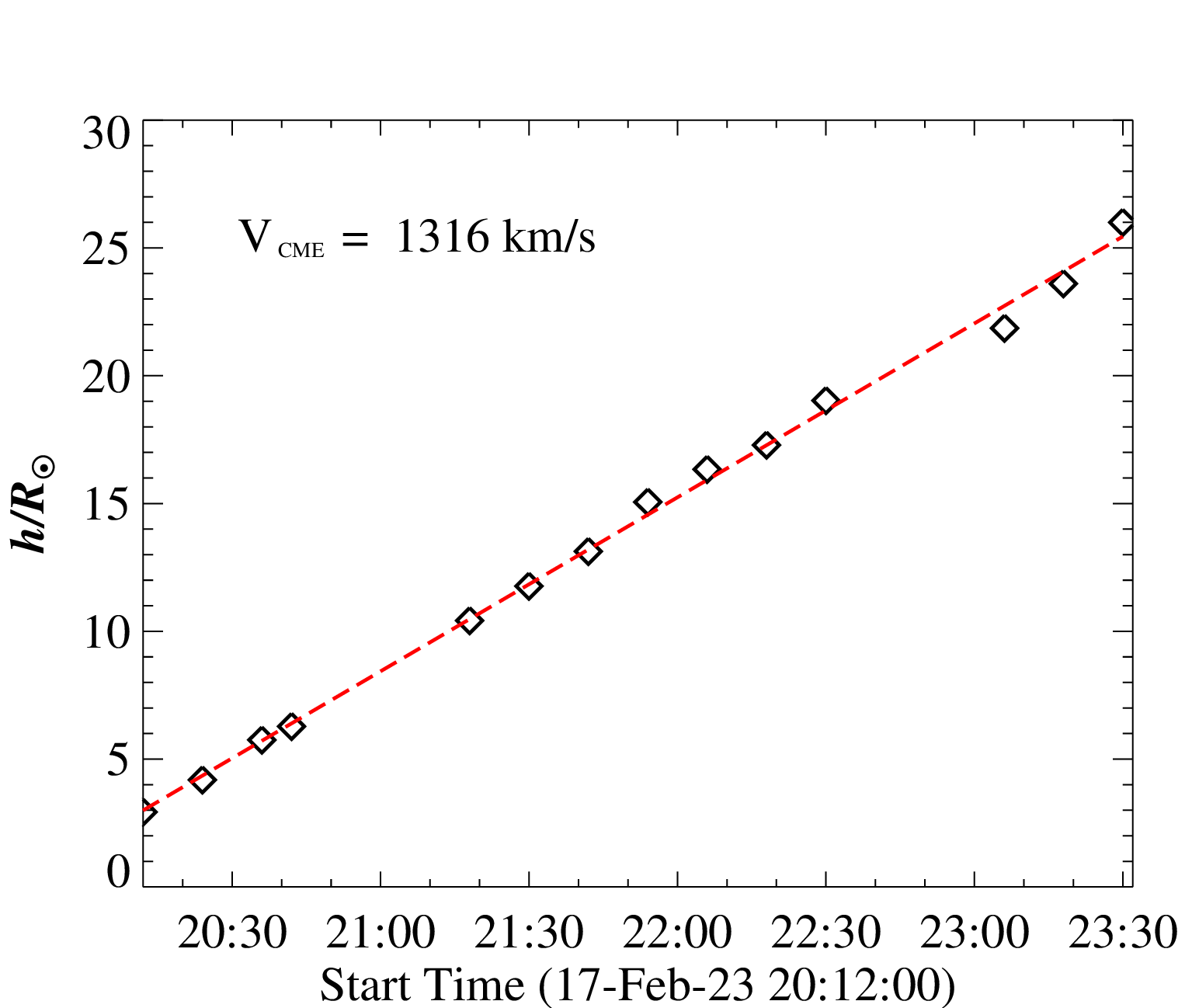}
    \caption{Height-time plot of the CME leading edge observed by SOHO/LASCO. 
    The linear speed ($\sim$1316 km s$^{-1}$) of the CME is labeled.}
    \label{fig7}
\end{figure}

The eruption of HC also generates a fast full-halo CME.
In Figure~\ref{fig6}, the top and bottom panels demonstrate running-difference images of the CME observed by LASCO-C2 and LASCO-C3, respectively.
The CME first appears at 20:12 UT and propagates in the same direction as the HC (PA$\sim$45$^{\circ}$). 
The CME presents a typical three-part structure, including a bright leading edge, a dark cavity, and a bright core \citep{ill85,zhou23}. 
The bright core originates from the HC in this case.
Figure~\ref{fig7} shows the height-time plot of the CME. A linear fitting results in an apparent speed of $\sim$1316 km s$^{-1}$ in the plane of the sky.
Assuming a radial evolution without deflection, the true speed of the CME reaches $\frac{1316}{\sin(67^{\circ})}\approx1520$ km s$^{-1}$.
Such a fast CME is sufficiently capable of driving a shock wave, which is pointed by yellow arrows in Figure~\ref{fig6}.

\begin{figure}
    \centering
    \includegraphics[width=1.0\columnwidth]{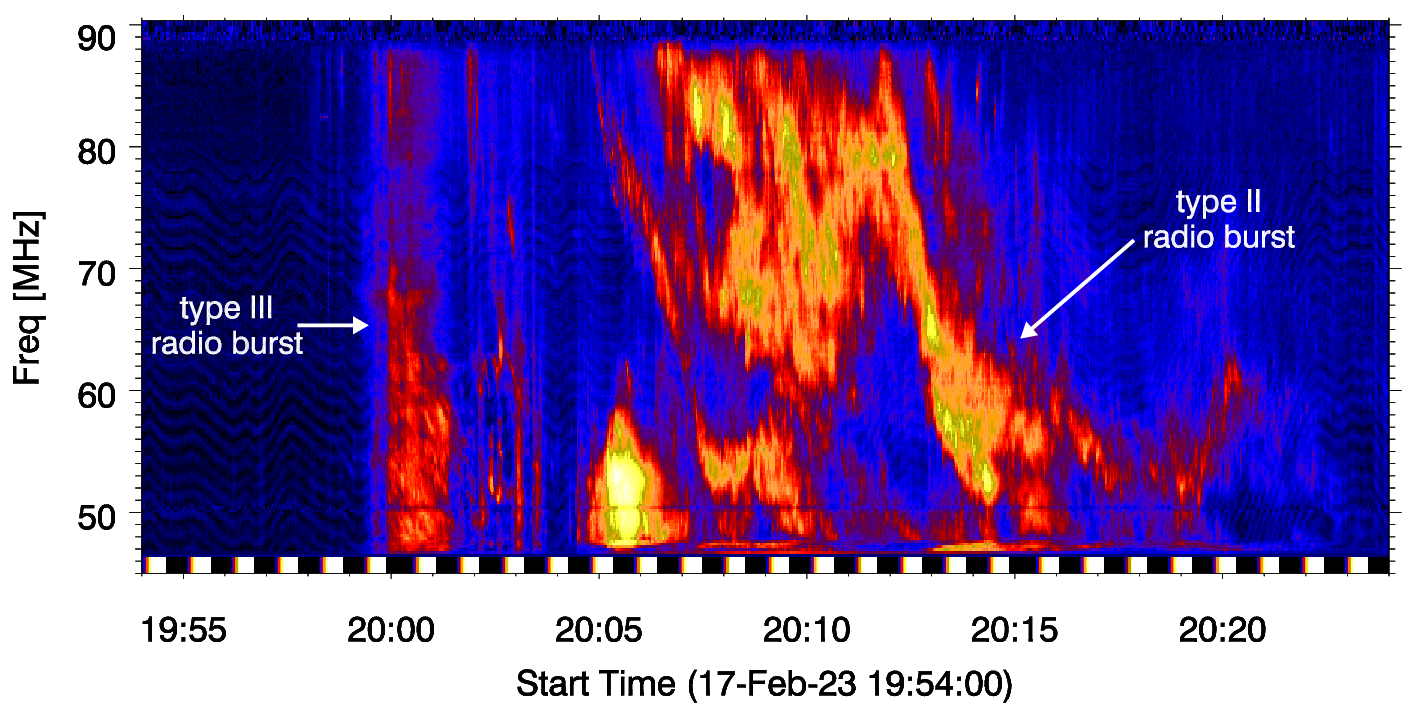}
    \caption{Radio dynamic spectra recorded by e-Callisto/Mexico-LANCE-A radio telescope during the flare. 
    The yellow arrows point to the type II and type III radio bursts.}
    \label{fig8}
\end{figure}

Figure~\ref{fig8} shows the radio dynamic spectra detected by e-Callisto/Mexico-LANCE-A radio telescope in the frequency range of 45$-$90 MHz during the flare.
The spectra are characterized by a type III burst around 20:00 UT and a following type II burst with a slower frequency drift rate than type III burst.
The type III radio burst is cotemporary with the flare impulsive phase as well as the fast rise of the HC \citep{zqm24}. 
The presence of type II radio burst is consistent with the CME-driven shock \citep{mor19,zqm22}.
Using MHD numerical simulations, \citet{xing24} investigated the initiation of a CME.
It is found that the slow rise is first triggered and driven by the developing hyperbolic flux tube (HFT) reconnection. 
Then, the slow rise continues as driven by the coupling of the HFT reconnection and the early development of torus instability (TI). 
The end of the slow rise, i.e., the onset of the impulsive acceleration (fast rise), is induced by the start of the fast MR (type III radio burst) coupled with the TI.

\subsection{EUV wave and filament oscillations} \label{sect:fo}
\begin{figure*}
    \centering
    \includegraphics[width=1.5\columnwidth]{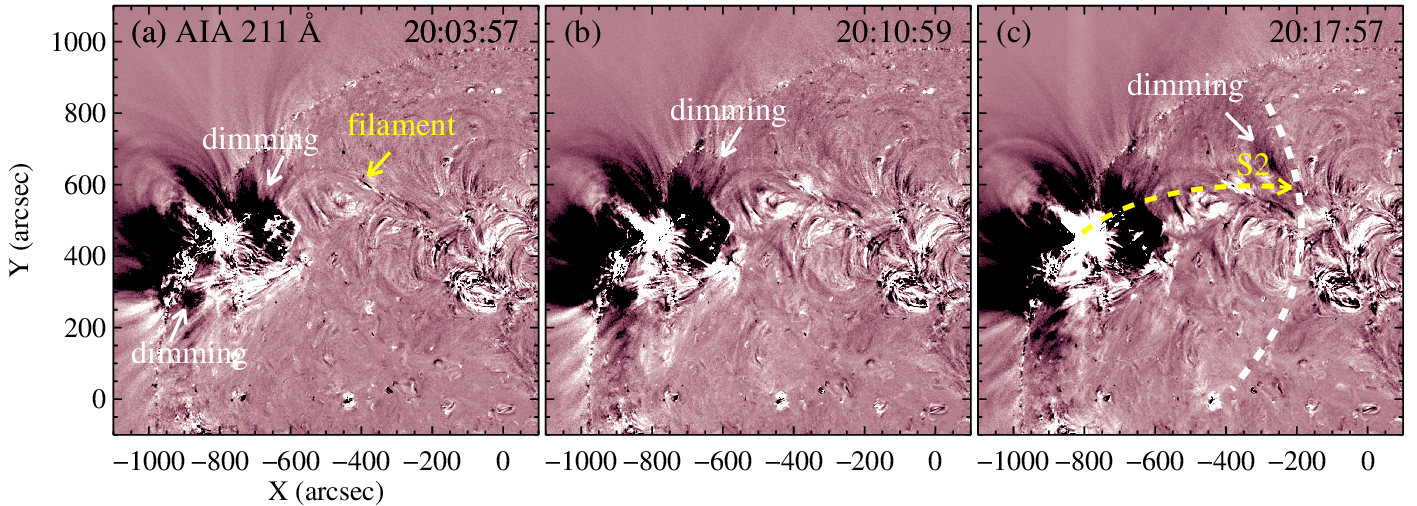}
    \caption{AIA 211 {\AA} base-difference images to illustrate the propagation of EUV wave. The white arrows points to the dark coronal dimmings.
    In panel (a), the yellow arrow points to the quiescent filament. In panel (c), the white dashed line denotes the EUV wave front.
    The yellow dashed line (S2) is used to calculate the speed of EUV wave.
    An online animation of AIA 211 {\AA} base-difference images is available as Supplementary material. 
    The $\sim$2 s animation covers from 20:03 UT to 20:18 UT.}
    \label{fig9}
\end{figure*}

\begin{figure}
    \centering
    \includegraphics[width=0.9\columnwidth]{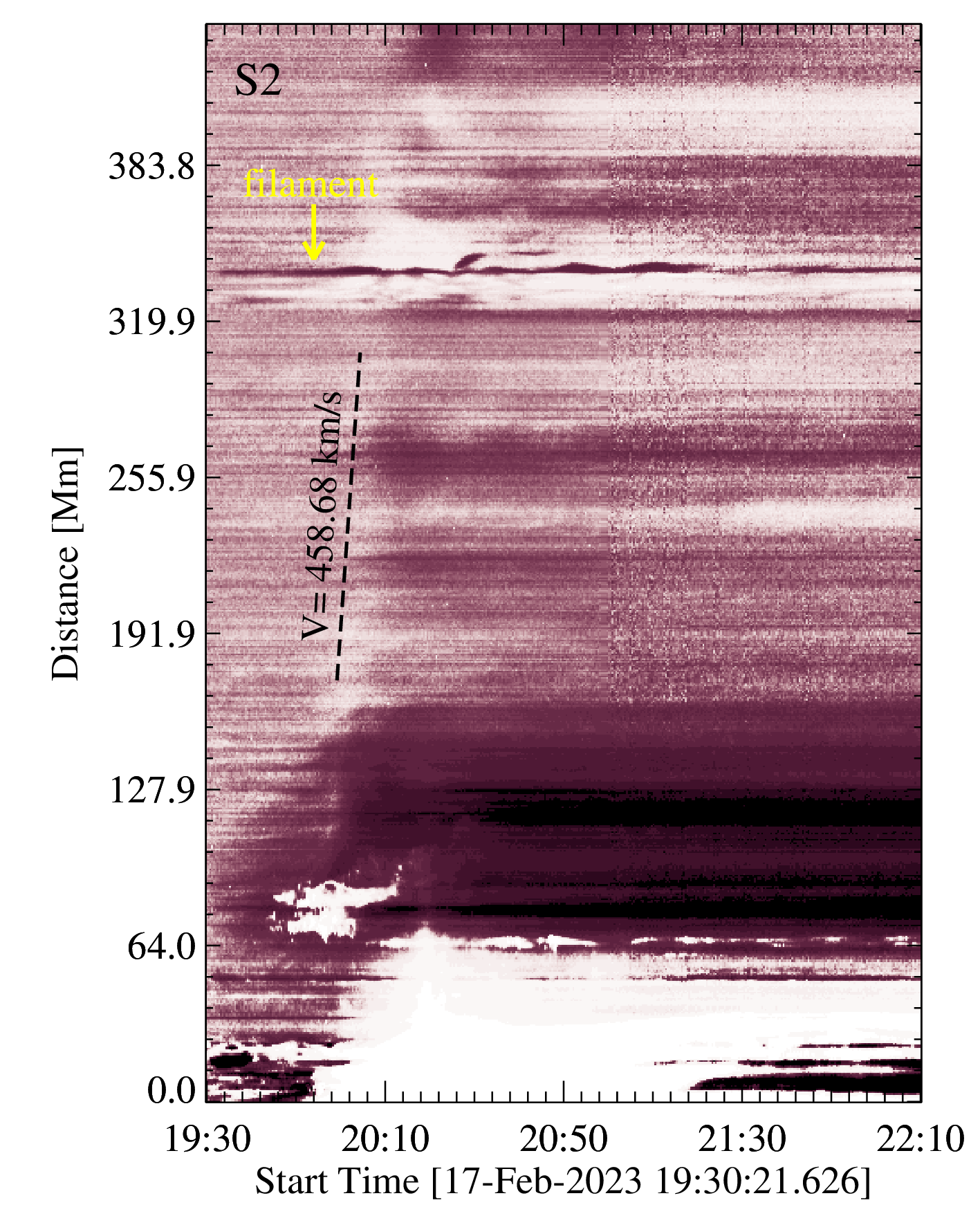}
    \caption{Time-distance diagram of S2 using AIA 211 {\AA} base-difference images.
    The speed of EUV wave (458.68 km s$^{-1}$) is labeled. The filament position along S2 is indicated by the yellow arrow.}
    \label{fig10}
\end{figure}

The fast CME drives a coronal EUV wave spreading outward from the flare site. 
Figure~\ref{fig9} shows three snapshots of AIA 211 {\AA} base-difference images to illustrate the propagation of EUV wave during 20:03$-$20:18 UT
(see also the online animation \textit{anim3.mp4}). In panel (a), the white arrows point to the dark coronal dimming regions, while the yellow arrow points to the filament.
In panel (c), the white dashed line denotes the EUV wave front. To calculate its speed, we select a curved slice (S2) with a total length of $\sim$441.4 Mm in panel (c).
Time-distance diagram of S2 in AIA 211 {\AA} is displayed in Figure~\ref{fig10}. The speed of the EUV wave is calculated to be $\sim$458.7 km s$^{-1}$.
The yellow arrow indicates the position of filament along S2. It is seen that the EUV wave arrives at the filament around 20:04 UT and induces oscillations.
After passing the filament, the wave continues to propagate.

\begin{figure*}
    \centering
    \includegraphics[width=1.8\columnwidth]{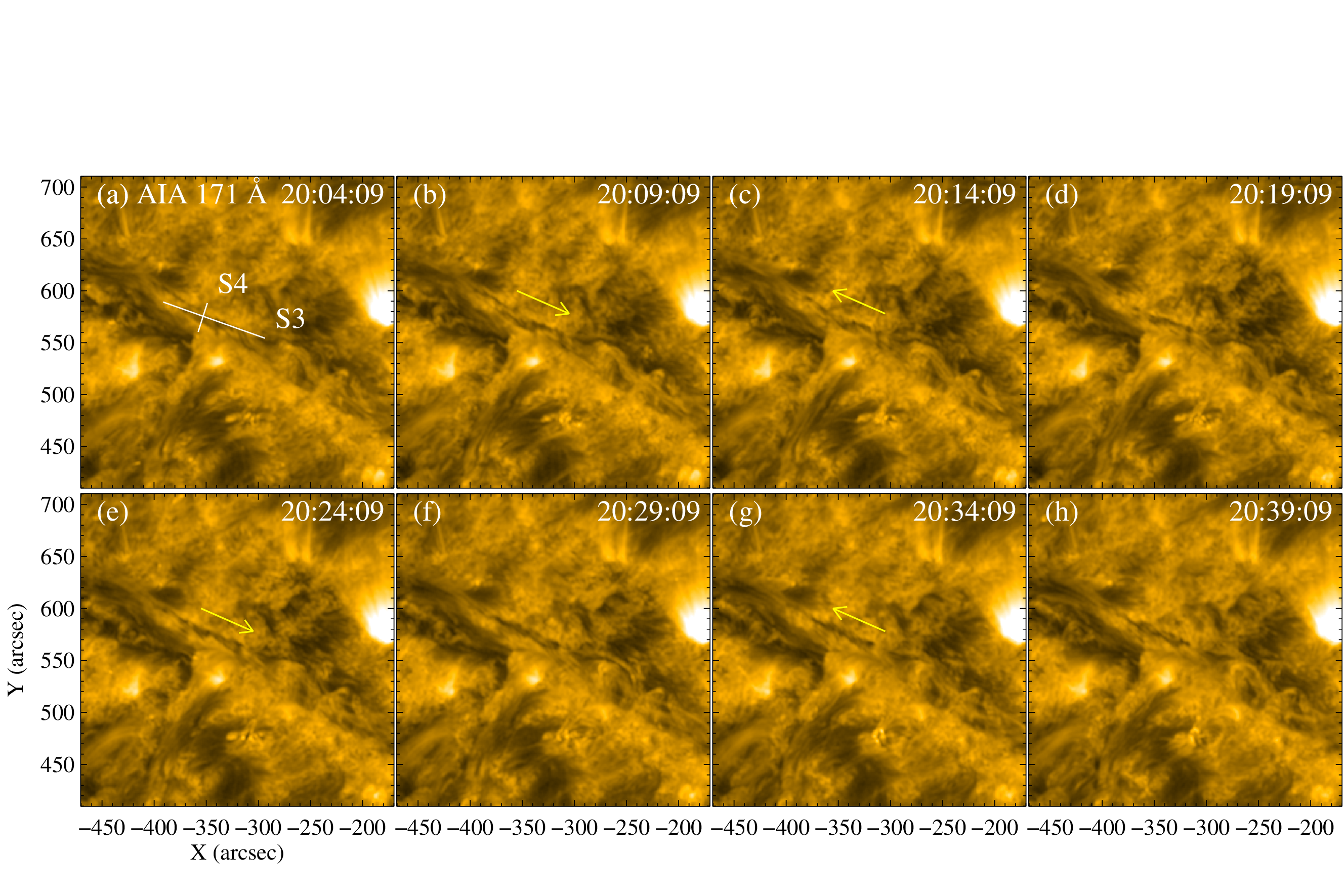}
    \caption{AIA 171 {\AA} images, showing LALOs of the filament. The yellow arrows illustrate the direction of filament movement.
    In panel (a), two straight slices (S3 and S4) are used to investigate LALOs and LATOs of the filament.
    An online animation of AIA 171 {\AA} images is available as Supplementary material. 
    The $\sim$12 s animation covers from 20:04 to 20:39 UT.}
    \label{fig11}
\end{figure*}

\begin{figure*}
    \centering
    \includegraphics[width=1.5\columnwidth]{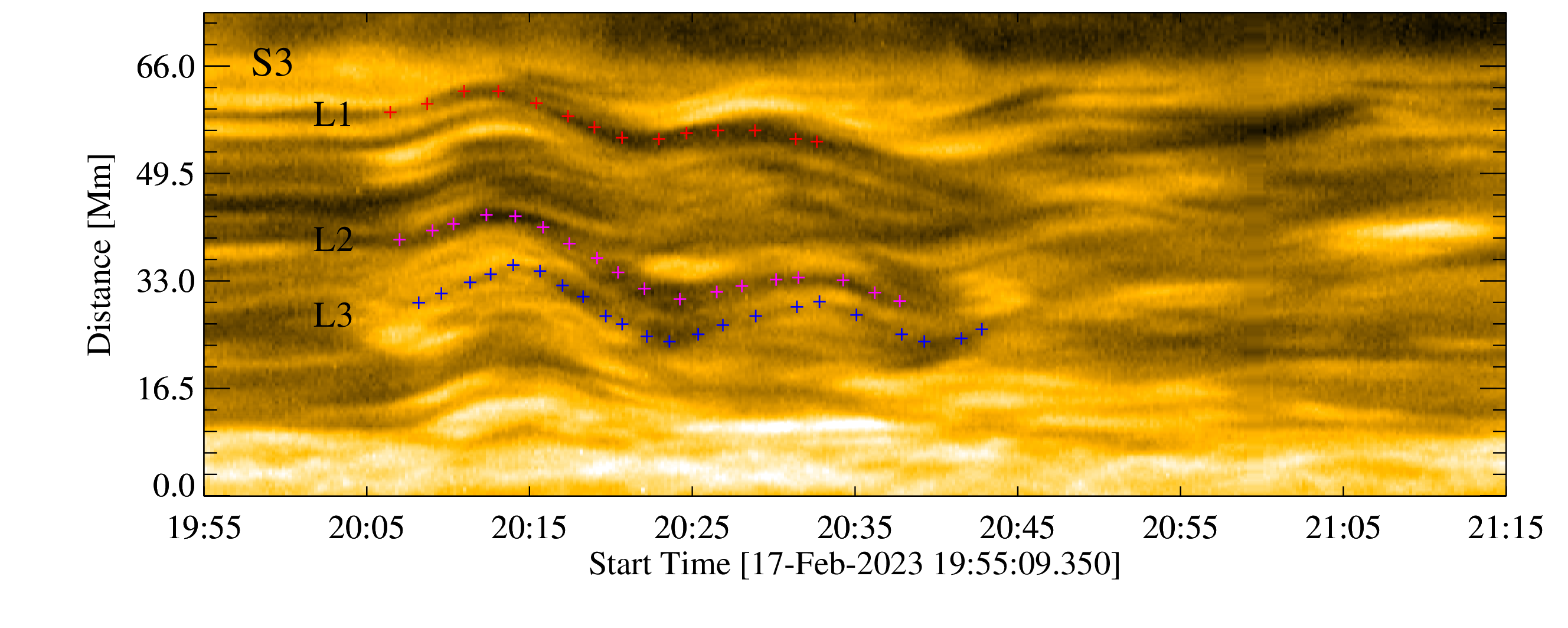}
    \caption{Time-distance diagram of S3 in AIA 171 {\AA}.
    $s=0$ and $s=74.3$ Mm represent the northeast and southwest endpoints of S3, respectively.
    Three oscillation patterns are selected and marked with red, magenta, and blue plus symbols.}
    \label{fig12}
\end{figure*}

Figure~\ref{fig11} shows eight snapshots of AIA 171 {\AA} images during 20:04$-$20:39 UT (see also the online animation \textit{anim4.mp4}). 
The yellow arrows indicate the directions of filament movement. After the impact of EUV wave, the filament first moves along the spine in the southwest direction (panel (b)).
Then it returns back and moves in the northeast direction (panel (c)). Such a longitudinal oscillation lasts for a few cycles before coming to a halt.
To investigate the longitudinal oscillation, we select a straight slice (S3) with a length of $\sim$74.3 Mm in panel (a).
Time-distance diagram of S3 in AIA 171 {\AA} is displayed in Figure~\ref{fig12}. 
The red, magenta, and blue plus symbols represent three oscillation patterns (L1, L2, and L3).
To derive the parameters of longitudinal oscillations, we perform curve fittings using the following function \citep{zqm24}:
\begin{equation} \label{eqn-4}
 	y(t)=A_0 \sin(\frac{2\pi t}{P}+\phi) e^{-\frac{t}{\tau}} +k t + y_0,
\end{equation}
where $A_0$, $\phi$, and $y_0$ are initial amplitude, phase, and position along S3, respectively. $P$ and $\tau$ are period and damping time of longitudinal oscillations.
$k$ denotes linear velocity of the filament.

\begin{figure}
    \centering
    \includegraphics[width=\columnwidth]{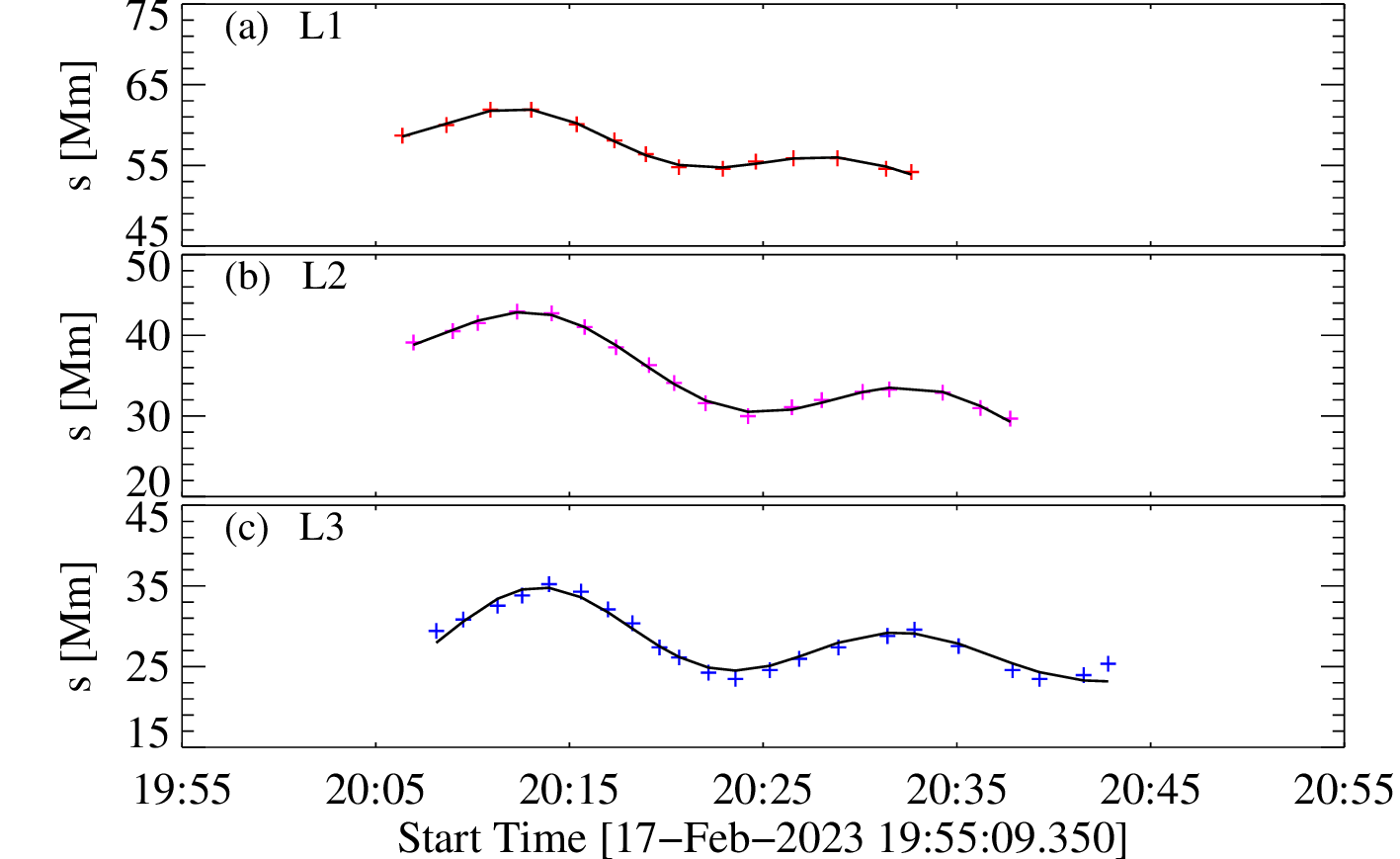}
    \caption{Three oscillation patterns (L1, L2, and L3) are drawn with red, magenta, and blue plus symbols, respectively.
    Results of curve fittings using Equation~\ref{eqn-4} are superposed with black solid lines.}
    \label{fig13}
\end{figure}

The results of curve fittings are plotted with black solid lines in Figure~\ref{fig13}. The corresponding parameters are listed in Table~\ref{tab-1}.
It is found that the filament threads start oscillating longitudinally after the impact of EUV wave at $\sim$20:05 UT and last for nearly two cycles. 
The three patterns are well fitted with Equation~\ref{eqn-4}. The initial amplitudes range from 3.2 to 6.3 Mm with an average value of $\sim$4.7 Mm.
The periods are between 985 and 1192 s with an average value of $\sim$1099 s.
The damping times are 1.5$-$4.0 times longer than the periods.
Negative values of $k$ indicate that the filament experiences a northeast drift as a whole at speeds of a few km s$^{-1}$.
The values of velocity amplitude ($v_0=\frac{2\pi A_0}{P}$) are listed in the last column of Table~\ref{tab-1}.
$v_0$ ranges from 20 to 35 km s$^{-1}$ with an average value of $\sim$26.5 km s$^{-1}$, consolidating that the longitudinal oscillations are typically large-amplitude.

\begin{table*}
	\centering
	\caption{Parameters of LALOs of the filament along S3.}
	\label{tab-1}
	\begin{tabular}{ccccccccc}
		\hline
		Pattern & $A_0$    & $P$   & $\phi$ & $\tau$ & $k$      & $y_0$ & $\frac{\tau}{P}$ & $v_0$   \\
                             & (Mm) & (s) & (rad)  & (s)    & (km s$^{-1}$) & (Mm)  &  $-$ & (km s$^{-1}$) \\
		\hline
	        L1   & 3.22 & 984.73  & 5.25 & 1840.11 & -5.09 & 61.35 & 1.87 & 20.57       \\
                L2   & 4.49 & 1192.41 & 5.66 & 4803.03 & -7.09 & 41.45 & 4.03  &  23.68    \\
                L3   & 6.31 & 1120.15 & 5.85 & 1637.74 & -2.76 & 30.63 & 1.46  & 35.37     \\
                \hline
                Avg. & 4.67 & 1099.10 & 5.59 & 2760.29 & -4.98 & 44.48 & 2.45  & 26.54      \\
		\hline
	\end{tabular}
\end{table*}

\begin{figure*}
    \centering
    \includegraphics[width=1.5\columnwidth]{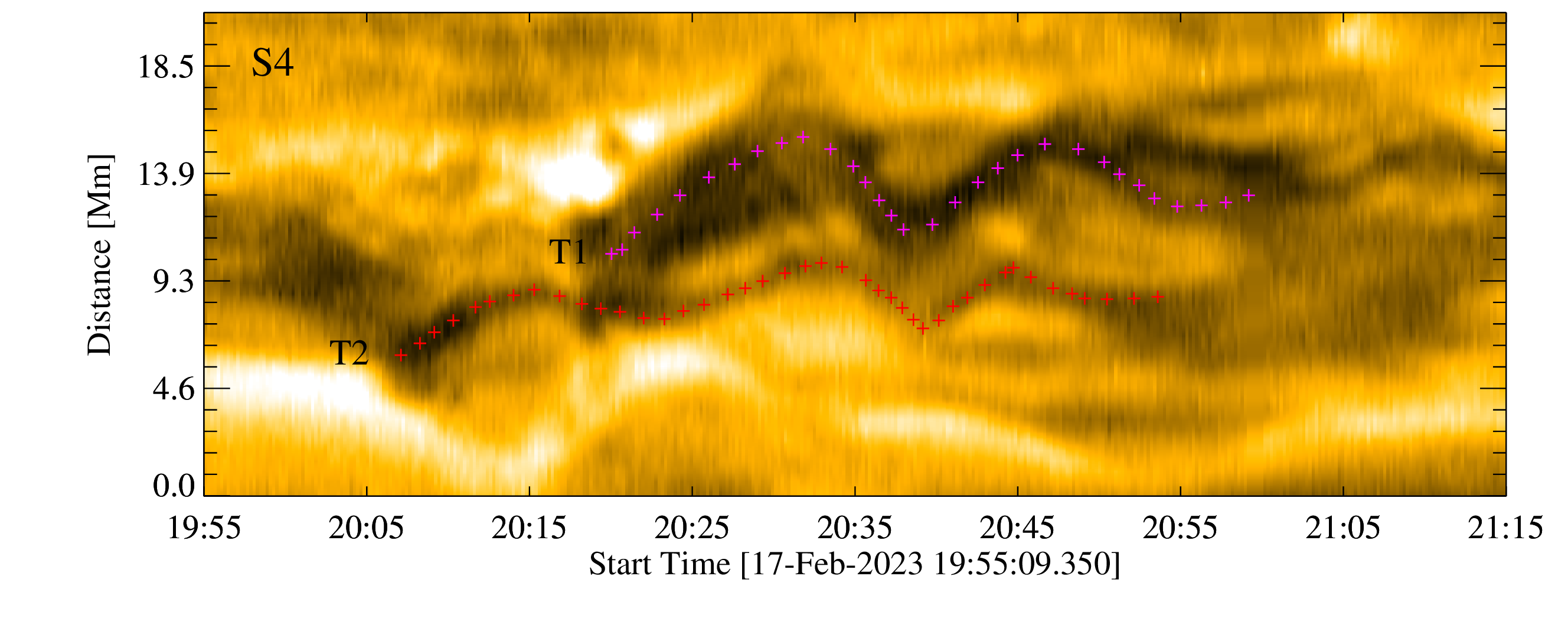}
    \caption{Time-distance diagram of S4 in AIA 171 {\AA}. $s=0$ and $s=20.8$ Mm denote the southeast and northwest endpoints of S4, respectively.
    Two oscillation patters (T1 and T2) are marked with magenta and red plus symbols, respectively.}
    \label{fig14}
\end{figure*}

\begin{figure}
    \centering
    \includegraphics[width=\columnwidth]{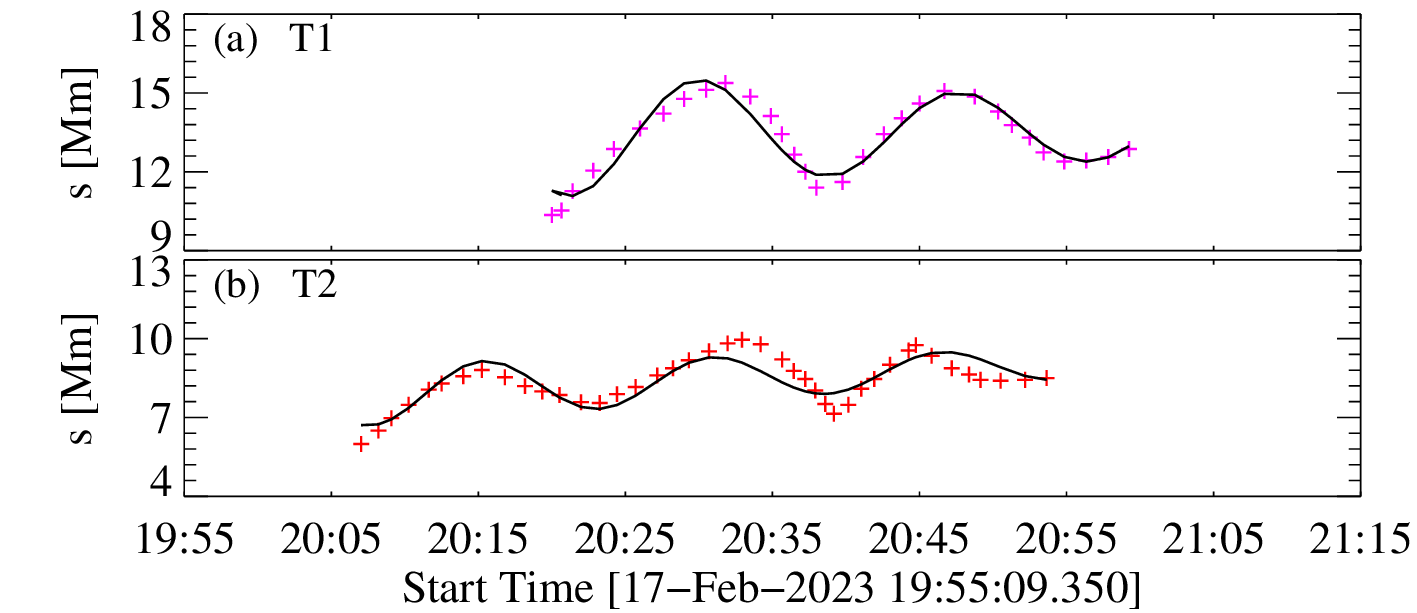}
    \caption{Two oscillation patterns (T1 and T2) are drawn with magenta and red plus symbols.
    Results of curve fittings using Equation~\ref{eqn-4} are superposed with black solid lines.}
    \label{fig15}
\end{figure}

It is revealed from Figure~\ref{fig4}(a) and Figure~\ref{fig9}(c) that the propagation of EUV wave along S2 has an included angle of $\sim$45$^{\circ}$ with the filament spine.
According to the cartoon model \citep{shen14a}, it is very likely that transverse oscillations are excited in the meantime.
In Figure~\ref{fig11}(a), we select a short slice (S4), which is perpendicular to S3 and has a length of $\sim$20.8 Mm.
Time-distance diagram of S4 in AIA 171 {\AA} is plotted in Figure~\ref{fig14}. 
Two transverse oscillation patters (T1 and T2) are marked with magenta and red plus symbols, respectively.
It is found that T2 starts oscillating at $\sim$20:06 UT and lasts for about three cycles, while T1 starts oscillating at $\sim$20:20 UT and lasts for two cycles.
Likewise, we perform curve fittings for T1 and T2 using Equation~\ref{eqn-4}.
In Figure~\ref{fig15}, T1 and T2 are drawn with magenta and red plus symbols in the top and bottom panels.
The results of fittings are superposed with black solid lines.
The corresponding parameters of LATOs are listed in Table~\ref{tab-2}.
The initial amplitude is between 1.2 and 2.4 Mm with an average value of $\sim$1.8 Mm, which is $\sim$2.6 times smaller than that of LALOs.
The period is between 932 and 1056 s with an average value of $\sim$994.4 s, which is shorter than that of LALOs.
The damping time is between 3297 and 3855 s with an average value of $\sim$3576 s, which is $\sim$1.3 times larger than that of LALOs.
The quality factor $q$ is larger as well, indicating that LATOs with smaller amplitudes attenuate slower and last longer than LALOs in this case.
The initial velocity of LATOs is between 8 and 14 km s$^{-1}$ with an average value of $\sim$11.2 km s$^{-1}$, which is $\sim$2.4 times slower than that of LALOs.
A positive value of $k$ indicates that the filament drifts gradually in the northwest direction during the oscillations.
It should be emphasized that we have explored different angles of the slices (S3 and S4) approximately following the apparent motion of the filament.
We found that the space-time diagrams do not change significantly, suggesting that there are indeed a mix of longitudinal and transverse oscillations 
induced by the EUV wave.


\begin{table*}
	\centering
	\caption{Parameters of LATOs of the filament along S4.}
	\label{tab-2}
	\begin{tabular}{ccccccccc}
		\hline
		Pattern & $A_0$    & $P$   & $\phi$ & $\tau$ & $k$      & $y_0$ & $\frac{\tau}{P}$ & $v_0$   \\
                             & (Mm) & (s) & (rad)  & (s)    & (km s$^{-1}$) & (Mm)  &   &  (km s$^{-1}$)     \\
		\hline
                T1   & 2.39 & 1056.34 & 4.47 & 3297.04 & 0.09 & 13.43 & 3.12 & 14.21      \\
                T2   & 1.22 & 932.43 & 4.44 & 3855.29  & 0.40 & 7.88 & 4.13 &  8.22     \\
                \hline
		Avg. & 1.81 & 994.39 & 4.45 & 3576.17 & 0.25 & 10.66 & 3.63 & 11.22       \\
		\hline
	\end{tabular}
\end{table*}

\section{Discussion} \label{sect:dis}
As mentioned in Section~\ref{sect:intro}, prominence seismology is an important tool to infer the magnetic field strength of dips supporting the dense prominences \citep{arr18}.
For LALOs, the predominant restoring force is the gravity of threads along the dips \citep{lk12,zqm12,zqm13,zhou14,luna16}, 
although magnetic pressure gradient force might play a role \citep{vrs07}. 
In the pendulum model \citep{lk12}, the period ($P_L$) is solely dependent on the curvature radius ($R$) of the dip (see Equation~\ref{eqn-1}).
Therefore,
\begin{equation} \label{eqn-5}
R\approx 0.025\times P_L^2,
\end{equation}
where $P_L$ is in unit of minute and $R$ is in unit of Mm.
According to the observed values of $P_L$ listed in Table~\ref{tab-1}, $R$ is estimated to be 6.7$-$9.9 Mm.
In addition, the minimum magnetic field strength ($B_h$) along the filament threads is \citep{luna14}:
\begin{equation} \label{eqn-6}
B_h \geq \frac{(17\pm9)}{60}P_{L},
\end{equation}
where $B_h$ is in unit of G. Using the same values of $P_L$, the minimum $B_h$ is in the range of 4.6$-$5.6 G.

\citet{rud16} derived the periods and damping times of longitudinal filament oscillations in a magnetic flux tube analytically.
In the first case when the tube has two straight parts and a curved dip in the center, the dimensional period $P_n$ consists of two parts:
\begin{equation} \label{eqn-7}
P_{n}=P_{\mathrm{shift}}+P_{g},
\end{equation}
where the first term $P_{\mathrm{shift}}$ introduces the period shift as a result of mass accretion.
The period shift increases with the mass accretion rate. 
The second term $P_{g}$, which exclusively depends on the curvature radius, is equivalent to Equation~\ref{eqn-1}.
The expression assumes that the thread fills only the dipped part of the flux tube and the initial displacement is small compared with the curvature radius.
There are undoubtedly limitation and uncertainty in the estimation of curvature radius using the observed period.
In the current study, the periods of LALO almost keep constant during the oscillations (see Figure~\ref{fig13}).
Hence, the period shift and mass accretion are negligible. However, the initial amplitudes are non-negligible compared with the curvature radius (see Table~\ref{tab-1}).
Therefore, the estimation of curvature radius of the threads using Equation~\ref{eqn-5} may have large uncertainties.
Numerical simulations are worthwhile to carry out a more precise assessment in the future \citep{lia25}.

For LATOs, the radial magnetic field strength ($B_r$) of the filament is expressed as \citep{hyd66,pint08}:
\begin{equation} \label{eqn-8}
 	B_r^2=\pi\rho r_0^2(4\pi^2P_{T}^{-2}+\tau_{T}^{-2}),
\end{equation}
where $r_0\approx3.0\times10^9$ cm is the height of the filament \citep{hyd66}, 
$N_{\mathrm H}=2\times10^{10}$ cm$^{-3}$ and $\rho\approx1.27N_{\mathrm H}m_{\mathrm H}$ are number density and mass density of the filament \citep{shen14b},
$P_T$ and $\tau_T$ represent the period and damping time of LATOs. 
In most cases, $\tau_T$ is a few times longer than $P_T$, the second term is two orders of magnitude lower than the first term in Equation~\ref{eqn-8},
meaning that $B_r$ is mainly dependent on $P_T$:
\begin{equation} \label{eqn-9}
 	B_r\approx \frac{2\pi r_{0}}{P_T}\sqrt{\pi\rho}.
\end{equation}
Using the observed values of $P_T$ and $\tau_T$ in Table~\ref{tab-2}, $B_r$ is in the range of 6.6$-$7.4 G.
Therefore, the total magnetic field strength of the filament reaches 8.0$-$9.3 G.
In Table~\ref{tab-3}, we compare the value of $B_r$ with previous works.
It is found that the estimated $B_r$ in quiescent filaments are close to each other ($<$10 G) and are significantly weaker than those in AR filaments \citep{est25}.

\begin{table*}
	\centering
	\caption{Comparison of $B_r$ (G) with previous works.}
	\label{tab-3}
	\begin{tabular}{cccccc}
		\hline
		Date & 2011/08/09 & 2016/03/21 & 2022/10/02 & 2014/09/01 & 2023/02/17  \\
		\hline
                $B_r$ & 8.1 & 7 & 6.5 & 5.4 & 6.6$-$7.4 \\
                \hline
		Ref. & \citet{shen14b} & \citet{shen17} & \citet{dai23} & \citet{zyj24} & this work \\
		\hline
	\end{tabular}
\end{table*}

Simultaneous LALOs and LATOs could be excited not only by EUV or shock waves \citep{shen14a}, but also by coronal jets.
\citet{zqm17a} investigated a coronal jet originating from AR 12373 and its interaction with a remote quiescent prominence 
after propagating in the northwest direction on 2015 June 29.
The transverse oscillation of the eastern part of prominence could be divided into two phases.
The initial amplitude increases from $\sim$4.5 to $\sim$11.3 Mm, while the initial velocity decreases from $\sim$20 to $\sim$10 km s$^{-1}$.
The period increases by a factor of $\sim$3.5. The western part of the prominence undergoes a transverse oscillation with a much smaller amplitude.
Meanwhile, the horizontal threads experience a longitudinal oscillation with significantly larger initial amplitude, velocity, and period.
\citet{tan23} reported simultaneous LALOs and LATOs in a quiescent filament driven by a two-sided-loop jet on 2011 November 29.
The LALO occurs at the bottom of a cavity with an amplitude and a period of $\sim$13 Mm and $\sim$1.2 hr.
The north part of the filament shows LATO with a much smaller amplitude ($\sim$3 Mm) and a shorter period ($\sim$0.33 hr).
The magnetic field strength of the filament is estimated to be 22$\pm$1 G.
\citet{luna21} performed 2.5D numerical simulations to explore the interaction between a jet and a filament channel supporting a prominence.
It is revealed that both LALOs and LATOs are excited by the jet. 
Interestingly, the prominence mass flows out of the dip and down to the chromosphere in case of a strong impact \citep{zqm13}.

In the past decade, extensive investigations of prominence oscillations have been carried out using numerical simulations \citep{zhou18,lia20}.
\citet{lia20} performed 2.5D numerical simulations of LAOs of prominences supported by flux ropes.
It is uncovered that external disturbance, such as a Moreton or EUV wave, perturbs the flux rope and excites LAOs of horizontal and vertical polarizations.
Besides, the periods of longitudinal and transverse oscillations show weak dependence on the shear angle of the magnetic structure and prominence density contrast.
In a follow-up work, \citet{lia23} found that the external triggering of prominence oscillations is very complex, which involves LAOs or SAOs of the longitudinal or transverse
polarizations or a mixture of both types. Sufficient energy is required to induce LAOs. 
\citet{dai23} studied the transverse oscillations of a quiescent filament excited by an EUV wave on 2022 October 2.
The minimum energy of the EUV wave is estimated to be 2.7$\times$10$^{20}$ J.
In addition, the orientation of the prominence axis with respect to the driving event plays an important role in determining the certain type of LAOs \citep{lia23}.
In the current study, the included angle between the incoming EUV wave and filament spine is $\sim$45$^{\circ}$, 
which is suitable to induce both types of LAOs as described in Section~\ref{sect:fo}.
Recently, \citet{lia25} explored the interaction between a fast EUV wave and a remote prominence. 
It is seen that the interaction generates both transverse and longitudinal oscillations of the prominence. 
Meanwhile, magnetic reconnection takes place at a null point below the prominence flux rope, which is triggered by the fast wave.
Their simulation shows the full consequences of remote eruptions on prominence dynamics as reported in our current work.

\section{Summary} \label{sect:sum}
In this paper, we carry out multiwavelength observations of simultaneous longitudinal and transverse oscillations of a quiescent filament 
excited by an EUV wave on 2023 February 17. The main results are summarized as follows:
\begin{enumerate}
 \item The EUV wave is driven by a fast CME as a result of a HC eruption, which also generates an X2.3 class flare in AR 13229 close to the eastern limb. 
 The EUV wave propagates westward at a speed of $\sim$459 km s$^{-1}$.
 After arriving at the filament $\sim$340.3 Mm away from the flare site, 
 the EUV wave induces large-amplitude longitudinal and transverse oscillations of the filament, which are mainly observed in AIA 171 {\AA}.
 \item The longitudinal oscillations last for nearly two cycles. The average initial amplitude, velocity, period, and damping time are 
 $\sim$4.7 Mm, $\sim$26.5 km s$^{-1}$, $\sim$1099.1 s, and $\sim$2760.3 s, respectively.
 The curvature radius and minimum horizontal magnetic field strength of the dips are estimated to be 6.7$-$9.9 Mm and 4.6$-$5.6 G.
 \item The transverse oscillations last for 2$-$3 cycles. The average initial amplitude, velocity, period, and damping time are
 $\sim$1.8 Mm, $\sim$11.2 km s$^{-1}$, $\sim$994.4 s, and $\sim$3576.2 s, respectively.
 The radial magnetic field strength of the dips are estimated to be 6.6$-$7.4 G.
\end{enumerate}

\section*{Acknowledgements}
The authors appreciate the reviewer for valuable comments and suggestions to improve the quality of this article.
SDO is a mission of NASA\rq{}s Living With a Star Program. AIA and HMI data are courtesy of the NASA/SDO science teams.
The CHASE mission is supported by China National Space Administration (CNSA).
SUTRI is a collaborative project conducted by the National Astronomical Observatories of CAS, 
Peking University, Tongji University, Xi'an Institute of Optics and Precision Mechanics of CAS and the Innovation Academy for Microsatellites of CAS.
The radio dynamic spectra from e-Callisto website are provided by the Institute for Data Science FHNW Brugg/Windisch, Switzerland.
This work is supported by the National Key R\&D Program of China 2021YFA1600500 (2021YFA1600502), 2022YFF0503003 (2022YFF0503000), 
NSFC under the grant numbers 12373065, 12403068, Natural Science Foundation of Jiangsu Province (BK20231510), 
and Project Supported by the Specialized Research Fund for State Key Laboratory of Solar Activity and Space Weather.
\section*{Data Availability}
The data underlying this article will be shared on reasonable request to the corresponding author.


\bsp	
\label{lastpage}
\end{document}